# Opportunities of Optical Spectrum for Future Wireless Communications

Mostafa Zaman Chowdhury, Moh. Khalid Hasan, Md. Shahjalal, and Yeong Min Jang
Dept. of Electronics Engineering, Kookmin University, Seoul, Korea
E-mail: mzaman@kookmin.ac.kr, khalidrahman45@ieee.org, mdshahjalal26@ieee.org, yjang@kookmin.ac.kr

*Abstract*— **The requirements in terms of service quality such as data rate, latency, power consumption, number of connectivity of future fifth-generation (5G) communication is very high. Moreover, in Internet of Things (IoT) requires massive connectivity. Optical wireless communication (OWC) technologies such as visible light communication, light fidelity, optical camera communication, and free space optical communication can effectively serve for the successful deployment of 5G and IoT. This paper clearly presents the contributions of OWC networks for 5G and IoT solutions.**

## I. Introduction

The era of communication is shifting from fourth-generation (4G) to fifth-generation (5G). It is expected that 5G will be fully deployed in 2020. Upcoming 5G will offer new services with very high quality of service (QoS). The main features of the 5G communication services will include ultra-high system capacity, ultra-low latency, ultra-high security, massive device connectivity, ultra-low energy consumption, and extremely high quality of experience (QoE) [1]–[4]. It will provide 1000 times capacity compared to 4G. User will achieve up to 10 Gbps data rate. The latency will be within only few millisecond. It will reduce more than 90% energy consumption compared to 4G networks. To support the massive connectivity and high capacity demand, it is expected that 5G communication will comprise ultra-dense heterogeneous networks. The mobile data volume per area will be 1000 times and the number of connected wireless devices will be 100 times higher compared to 4G wireless networks. To achieve the goal, 5G networks must have the capability to support very high user data rate, very low energy consumption, and negligible end-to-end delay.

A massive number of end user devices or sensors are connected in Internet of Things (IoT). To envision the idea of Internet of Thing (IoT), the amount of connecting end user physical device with the internet is growing exponentially [5]. The end users communicate with each other without any human intervention. Therefore, IoT generates huge volume of data. 5G will connect a huge and more diverse set of devices to support IoT. The most important features of the IoT are the connection of massive devices and very low power consumption.

Currently, radio frequency (RF) is widely used for different wireless applications. To meet the demand of 5G features and IoT paradigm, only RF based wireless communication technologies are not sufficient. It is predicted that the demand for the massive connectivity of 5G and IoT data traffic will not be met by only using RF based wireless technologies. Therefore, researchers are finding new spectrum to fulfill the exponentially growing demands. International Telecommunication Union (ITU) proposed 11 new candidate bands for 5G communication [5]. Very large optical band is considered to be a promising solution for the development of high density and high capacity 5G and IoT networks. The wireless connectivity based on the optical spectrum is termed as optical wireless communication (OWC) [6], [7]. In comparison to RF-based networks, OWC-based network technologies offer unique advantages such as high data rate, low latency, high security, and low energy consumption. Communication distances ranging from a few nm to more than 10,000 km is possible using OWC systems [7]. The main technologies of OWC networks include visible light communication (VLC) [8]–[10] light fidelity (LiFi) [11]–[13], optical camera communication (OCC) [14]–[17], and free space optical (FSO) [18] communication. Each of these technologies has individual features. Indoors as well as outdoors services are offered by different OWC networks. These OWC technologies can play a vital role for achieving the goals of 5G and IoT. This paper explains the connection of OWC networks with 5G and IoT. We provide detail 5G and IoT solutions using different OWC networks.

## II. OWC Solutions for 5G and IoT

The RF band lies between 3 kHz and 300 GHz of the electromagnetic spectrum [7]. However, because of favorable communication properties, below 10 GHz is widely used by existing wireless technologies. This band is almost exhausted, and not able to provide 5G and IoT demands. Moreover, this band is strictly regulated by local and international authorities. The OWC is an excellent solution for high data rate communication. OWC has a wide range of applications [12]-[26], [29]-[41]. It can serve the demand of high data rate 5G services and massive connectivity of IoT applications. Various types of communications such as machine-to-machine, device-to-device, chip-to-chip, vehicle-to-vehicle, vehicle-to-infrastructure, infrastructure-to-vehicle, point-to-point, and point-to-multipoint can be accomplished using different OWC systems [7], [15]. The co-existence of RF and OWC networks can solve most of the limitations of RF based systems. This section briefly explains how OWC networks can provide effective solutions for 5G and IoT deployment.

*Ultra-high system capacity*: This is one of the key requirements for 5G networks. Also, to support massive IoT connectivity, high-speed network connectivity is needed. VLC as well LiFi can support very data rate services. Moreover, LiFi can support a complete network system i.e. point-to-multipoint, multipoint-to-point, and bidirectional communications as like as WiFi. A data rate of 100 Gbps was already confirmed using VLC [9]. FSO can also support high data rate services at indoor as well as outdoor. Outdoor remote high speed connectivity is possible using FSO network. OWC based on ultraviolet band can provide high-data rate non-line-of-sight communications. Hence, OWC networks are good complementary solution for 5G and IoT.

*Massive device connectivity:* OWC can play a vital role for providing massive connectivity. The usage of LEDs for different applications is increasing exponentially because of their low price, low energy consumption, and longer life span. Hence, OWC technology is able to provide enormous number of connection through low power LEDs to achieve the goals of 5G and IoT.

*Ultra-low latency:* The communication using optical band is much faster than that using RF bands, as the propagation is very fast in optical based communication systems. The processing time in optical system is also very fast. Hence, OWC systems take negligible end-to-end delay. Hence, OWC based network technologies can provide services with negligible latency.

*Ultra-high security:* OWC technologies can provide secure communication as required by 5G and IoT. The OWC signal cannot penetrate obstacle. Hence, outside people cannot hack the information. Especially for health purposes, the information can be exchanged with highly secured manner.

*Ultra-low energy consumption*: Huge researches are performing the researches around the world to reduce the power consumption by LEDs. OWC technologies consume very low power that is also one of the key requirements of 5G [1]. Present LEDs consume very small energy. Moreover, LEDs can be used for illumination as well as communication. Therefore, additional energy is not consumed by LED transmitter if it is used for illumination as well. Compared to RF sensors, LED sensors consume very small energy. Hence, the OWC based communication can save lot of energy that is an important requirement for 5G and IoT deployments.

*Extremely high quality of experience***:** OWC networks provide high throughput, low latency, and high security for indoor users. Hence, the indoor users can achieve very high level of QoE that is one of the most important features of 5G networks. As a result of dense OWC deployment, the call blocking probability [19 ] is reduced to negligible.

*Reliable network:* Reliable connectivity is an important criteria for 5G and IoT. Indoor optical wireless systems can provide reliable connectivity for users. Even, for outdoor scenarios, OCC, a sub-system of OWC, can provide non-interference communication and a high signal-to-noise ratio. Moreover, stable performance is achievable even when the communication distance increases. The outdoor

*Small cell networks***:** One of the most important characteristics of 5G communication networks is the deployment of highly dense small cell networks. The indoor VLC and LiFi create very highly dense small cells. Each of the networks under one light source is considered as a small cell. Their coverage area is very small. Even hundreds of VLC/LiFi based small cells can be found in a big room. Hence, OWC networks can fulfill the criteria for 5G networks. Fig. 1 shows example of OWC based highly dense small cell networks.

*Highly dense networks*: Highly dense deployment of networks is another important characteristic of 5G networks. The goal of 5G network can be achieved by deploying highly dense small cell networks. To meet the demands of future communication, there will be multi-tier networks e.g., satellite, macrocell, and small cells. The small cells contain RF femtocell as well as optical VLC, LiFi, and OCC networks. Each and every room can contain many optical small cells along with RF femtocells. Hence, deployment of OWC networks will help to deploy highly dense small cells. Fig. 1 shows that the OWC based small cell networks along with RF femtocells create highly dense networks deployment.

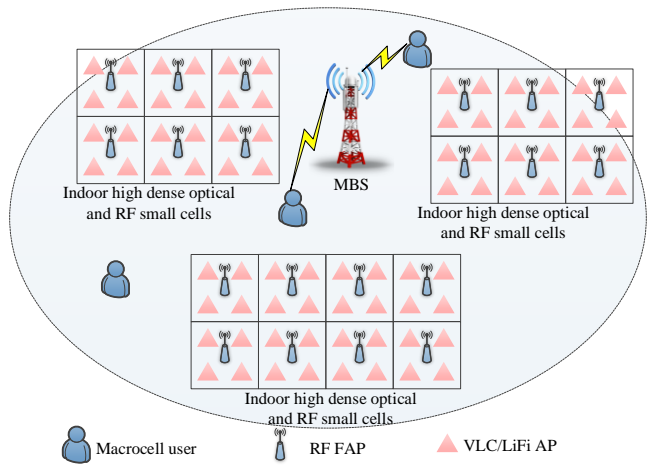

**Fig. 1.** A scenario of heterogeneous multi-tier networks containing a RF macrocell, many RF femtocells, and huge number of optical small cells.

*Convergence of heterogeneous networks:* RF based technologies has limitations in providing high data rate services and hence, OWC technologies became very promising solution to support high data rate 5G communication and massive connectivity of IoT. However, each of different RF and optical wireless technologies has limitations and advantages. The coexistence of heterogeneous networks can overcome the limitations. Hence, huge number of OWC networks can play a vital role in coexisting OWC/RF networks to mitigate limitations and to bring proper solution [10]. The multi-tier heterogeneous networks are very effective for load balancing. A huge number of users are served by indoor optical wireless networks. As a result, outdoor expensive and comparatively low capacity macrocell network can provide better services to outdoor users. Fig. 1 shows an example of multi-tier architecture consisting of macrocell/RF small cells/optical small cells.

*High capacity back-haul network:* High capacity backhaul is very important criteria for deploying high data rate communication networks. Without high capacity backhaul network, it is not possible to establish communication between access networks and core network. Hence, high data rate 5G systems must require very high capacity backhaul connectivity. Along with wired optical fiber networks, optical wireless networks such as FSO, LiFi, and VLC can effectively solve this issue. Specially, FSO can establish very high capacity and very long range outdoor backhaul link. Fig. 2 shows the example of high capacity backhaul connectivity using FSO communication. High capacity backhaul connectivity are established with the macrocellar base stations (MBSs) at remote island, remote hill, and remote city.

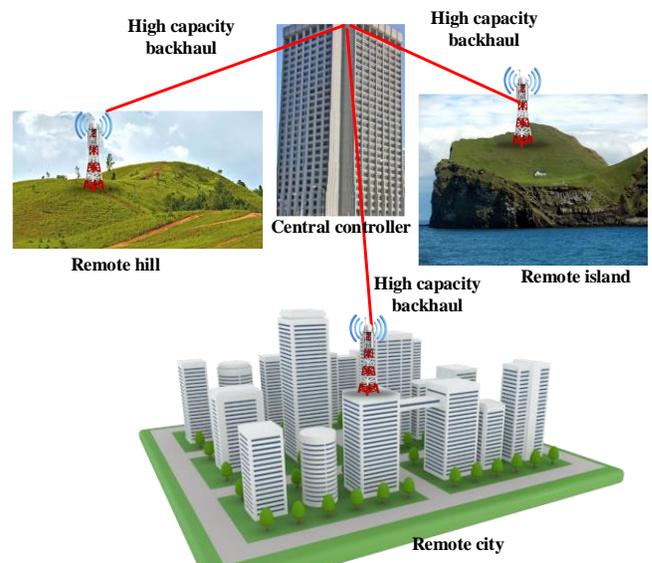

**Fig. 2.** High capacity backhaul connectivity for remote hill, remote island, and remote city.

### III. Conclusions

The 5G telecom networks will hit the market by 2020. The huge data produce by massive connectivity of IoT can be supported by 5G networks. The QoS requirements of 5G services are very high. Only RF based technologies are not able to fulfill this high demand. Hence, the complementary presence of optical based wireless technologies can effectively solve the limitations of RF based networks. This paper provides a detail observation regarding how OWC networks can help to achieve the goals of 5G and IoT paradigm.

### Acknowledgement

This research was supported by the MSIT (Ministry of Science and ICT), Korea, under the ITRC (Information Technology Research Center) support program (IITP-2018-0-01396) supervised by the IITP (Institute for Information & communications Technology Promotion) and the Korea Research Fellowship Program (2016H1D3A1938180).